\documentclass[preprint,showpacs,pra,aps,superscriptaddress]{revtex4}

\usepackage{graphicx}
\usepackage{dcolumn}
\usepackage{bm}
\usepackage{color}

\begin{document}


\title{Gigahertz decoy quantum key distribution with 1~Mbit/s secure key rate}

\author{A. R. Dixon}
\affiliation{Toshiba Research Europe Ltd, Cambridge Research
Laboratory, 208 Cambridge Science Park, Milton Road, Cambridge, CB4
0GZ, UK}

\affiliation{Cavendish Laboratory, University of Cambridge, J. J. Thomson Avenue, Cambridge CB3 0HE, UK}

\author{Z. L. Yuan}
\email{zhiliang.yuan@crl.toshiba.co.uk}
\affiliation{Toshiba Research Europe Ltd, Cambridge Research
Laboratory, 208 Cambridge Science Park, Milton Road, Cambridge, CB4
0GZ, UK}

\author {J. F. Dynes}
\affiliation{Toshiba Research Europe Ltd, Cambridge Research
Laboratory, 208 Cambridge Science Park, Milton Road, Cambridge, CB4
0GZ, UK}

\author {A. W. Sharpe}
\affiliation{Toshiba Research Europe Ltd, Cambridge Research
Laboratory, 208 Cambridge Science Park, Milton Road, Cambridge, CB4
0GZ, UK}

\author {A. J. Shields}

\affiliation{Toshiba Research Europe Ltd, Cambridge Research
Laboratory, 208 Cambridge Science Park, Milton Road, Cambridge, CB4
0GZ, UK}

\date{\today}

\begin{abstract}

We report the first gigahertz clocked decoy-protocol quantum key distribution (QKD). Record key rates have been achieved thanks to the use of self-differencing InGaAs avalanche photodiodes designed specifically for high speed single photon detection.  The system is characterized with a secure key rate of 1.02~Mbit/s for a fiber distance of 20~km and 10.1~kbit/s for 100~km. As the present advance relies upon compact non-cryogenic detectors, it opens the door towards practical and low cost QKD systems to secure broadband communication in future.

\end{abstract}


\maketitle

\section{Introduction}

Quantum key distribution (QKD) \cite{bennett84, bennett92,townsend93,gisin02,gottesman04} promises a method for distributing digital keys with unconditional security. With a sufficiently high key rate, QKD can reach its ultimate goal to allow information-theoretically secret communication using the one-time pad cipher, which has been ruled impractical due to the long-standing key distribution problem \cite{stallings}. Since its proposal \cite{bennett84} in 1984 and the subsequent proof-of-principle demonstration over a free space distance of 32~cm with a system clock rate of 200~Hz in 1991 \cite{bennett92}, experimental QKD has advanced significantly in both the distribution distance \cite{gobby04,takesue07} ($>$100 km) and the system clock rate \cite{takesue07, gordon05, yuan08} ($\geq$1~GHz). It has already emerged from laboratories for field tests in telecom network environments \cite{secoqc}.

Security is often a victim of fast-paced advances. In high speed QKD demonstrations, the security is commonly compromised due to the use of an attenuated laser source, which inevitably emits pulses containing more than one photon, instead of the single photon source of the original proposal. An eavesdropper (Eve) can launch a photon number splitting (PNS) attack \cite{dusek99,brassard00} on these multi-photon pulses to gain information without introducing errors in the distributed key.  Fortunately, the security loophole due to the multi-photon pulses can be overcome using the practical decoy protocol, developed independently by Lo \textit{et al.} \cite{lo05,ma05} and Wang \cite{wang05} based on the original idea by Hwang \cite{hwang03}. By replacing a subset of the signal pulses with less intense ``decoy" pulses, Alice and Bob can detect an attempted PNS attack through monitoring the quantum channel transmittances. As a simple modification to the original BB84 protocol, the decoy protocol provides a robust solution to prevent PNS attacks using only practical optical components. The decoy protocol promises long distance operation because its key rate is linearly dependent on the channel loss, as for BB84 implemented with an ideal single photon source, compared to the quadratic dependence of non-decoy BB84. Other protocols, such as coherent-one-way \cite{stucki05} and differential phase shift \cite{takesue07,inoue02}, are also effective against the PNS attack, but their unconditional security remains an open question \cite{scarani08}.

Decoy QKD has been demonstrated as a viable solution to achieve secure key distribution across fiber links \cite{yuan07a,rosenberg07,peng07,dynes07}. However, the secure key rate, limited to 10~kbit/s for a fiber distance of 20 km \cite{dynes07} and $\sim$10~bit/s for 100 km \cite{rosenberg07}, is insufficient for practical use. The limitation arises from the difficulty in counting frequently arriving single photons at a telecom wavelength. Recently, we demonstrated a simple self-differencing (SD) circuit \cite{yuan07b} that enables InGaAs avalanche photodiode (APD) to be used for high speed single photon detection.  Previously, we have successfully demonstrated the feasibility of using SD-APDs in a gigahertz QKD system, achieving multi-Mbit/s sifted key rate using the standard BB84 protocol \cite{yuan08}. In the present paper, we build upon that first result by incorporating decoy pulses into the standard BB84 protocol. This is the first report of gigahertz QKD with a protocol that is unconditionally secure against all type of attacks on the transmitted photons, including the PNS attack. Thanks to the high efficiency of the SD InGaAs APDs, the secure key rate has been elevated by two orders of magnitude over previous best values, and exceeds 1~Mbit/s over a fiber distance of 20~km for the first time. At 100~km, the secure key rate is determined to be 10.1 kbit/s.

\section{Decoy theory}

In a decoy QKD system, Alice replaces a subset of signal pulses (average intensity $\mu$ photons/pulse) randomly with weaker decoy pulses (average intensities $\nu_1$, $\nu_2$, $\cdot\cdot\cdot$). Individually, a decoy pulse is indistinguishable from a signal pulse.  If Eve carries out a PNS attack, by preferentially transmitting multi-photon pulses while blocking single photon pulses, she will inevitably transmit a lower fraction of the decoy pulses as they statistically contain fewer multi-photon pulses.  By calculating the transmission and error rates for signal and decoy pulses separately, Alice and Bob can place a bound on the information gained by Eve; information which can then be removed in the privacy amplification stage.

In BB84, only pulses containing exactly one photon contribute to secure key distribution, and the characteristics of these pulses can be rigorously bounded using the decoy protocol.  In a 3-state decoy protocol \cite{ma05}, where one signal state ($\mu$) and two decoy states ( $\nu_1$,  $\nu_2$) are used, one can obtain a lower bound for the single photon transmittance, $Q_1$, which represents the joint probability that a signal pulse contains only one photon and the photon is detected by Bob, using
\begin{equation}
Q_1 \geq Q_1^{L}=\frac{\mu^2e^{-\mu}}{\mu\nu_1-\mu\nu_2-\nu_1^2+\nu_2^2}[Q_{\nu_1}e^{\nu_1}-Q_{\nu_2}e^{\nu_2}-\frac{\nu_1^2-\nu_2^2}{\mu^2}(Q_{\mu}e^{\mu}-Y_0^L)].
\end{equation}
The upper bound for the single photon quantum bit error rate (QBER), $\varepsilon_1$, \textit{i.e.}, the QBER for those single photon pulses that are detected by Bob,  can be obtained using
\begin{equation}
\varepsilon_1\leq \varepsilon_1^U=\frac{\varepsilon_{\mu}Q_{\mu}e^{\mu}-\frac{1}{2}Y_0^L}{Q_1^Le^{\mu}},
\end{equation}			
where $\mu$, $\nu_1$ and $\nu_2$ are the average photon fluxes for signal and decoy pulses,  $Q_x$ ($x=\mu,\nu_1, \nu_2$) is the transmission probability of a signal or decoy pulse, and $\varepsilon_\mu$ is the QBER for signal state.  $Y_0^L$ is the lower bound for the count probability for an empty pulse (0 photon number), and can be obtained using
\begin{equation}
Y_0\geq Y_0^L = \frac{\nu_1Q_{\nu_2}e^{\nu_2}-\nu_2Q_{\nu_1}e^{\nu_1}}{\nu_1-\nu_2}
\end{equation}
With $Q_1^L$  and $\varepsilon_1^U$  available, the secure key rate can be determined using
\begin{equation}
R_{Secure}=\frac{1}{2}N_{\mu}\{-Q_{\mu}f_{EC}H_2(\varepsilon_{\mu})+Q_1^L[1-H_2(\varepsilon_1^U)]\}/t
\end{equation}
where  $N_{\mu}$ is the total number of the signal pulses sent by Alice, and $t$ is the QKD session time. $f_{EC}$ is the error correction efficiency and $H(x)=-x\log_2(x)-(1-x)\log_2(1-x)$ is the binary Shannon entropy. We use $f_{EC}=1.10$ in this paper based on previous results of our error correction module \cite{yuan07a}.

\section{Experimental setup}

Figure~1 shows a schematic diagram of the decoy QKD setup.  Photons are generated by a 1.55~$\mu$m pulsed laser operating at 1.036~GHz. The three intensity levels required for signal, decoy and near vacuum pulses are created using a fiber-optic intensity modulator. The quantum information is transmitted in the phase of the photons, with the phase encoding/decoding optics based on an asymmetric fiber Mach-Zender interferometer (AMZI). Off-the-shelf optical components were used to construct the AMZIs. The AMZI pair have matching time differences, of 440 ps in this case, to ensure that single photon interference takes place at the exit of the second AZMI.
Fine phase control is achieved through a fiber stretcher (not shown in Fig.~1) placed in one of Bob's AMZI arms \cite{gobby04,yuan05}. The AMZIs are carefully insulated from the environment to achieve sufficient stability to allow QKD to take place. The optical pulses are strongly attenuated to the desired levels before leaving the sender's apparatus, with the intensity level of the pulses monitored using a beam-splitter and optical power meter.  Dispersion shifted single mode fiber is used as the quantum channel, with an attenuation loss of 0.20~dB/km.  Note that standard single mode fiber SMF-28 can also be used as the quantum channel, as previously reported \cite{yuan08}, but fiber chromatic dispersion must be carefully compensated for long fibers ($>65$~km). A polarization controller at Bob's side recovers the polarization alignment before his AMZI.  Each photon detection event is recorded with a unique time-stamp by time-tagging electronics, and these events are used by Alice and Bob to generate a secure key.

\section{Self-diferencing I\lowercase{n}G\lowercase{a}A\lowercase{s} avalanche photodiodes}
InGaAs APDs are used for single photon detection at a clock rate of 1.036~GHz, cooled to --30~$^\circ$C using a thermoelectric cooler. We operate these devices in SD mode \cite{yuan07b}, as shown in Fig. 2(a). The raw APD output is processed by an inserted SD circuit to remove the periodic capacitive response before detection by the discrimination electronics.  After the SD avalanche signals are clearly identifiable from the background noise, as shown in Fig.~2(b).  The smallest separation between avalanches is 2~ns, which corresponds to 2 clock periods as expected for a perfect SD, suggesting an ultrashort detector dead time. Such a short dead time is three orders of magnitude less than typical values for conventional gated mode InGaAs APDs. We attribute this to the reduced avalanche charge, being 0.17~pC per avalanche, as inferred from the photo-current dependence on photon count in the linear regime. The detector biases were set to achieve a detection efficiency of 10\% with a dark count probability of $6.8\times10^{-6}$ per gate.

The SD APD has excellent timing performance, as shown in Fig.~2(c). The detector is biased active for  a 0.48-ns gate in each clock cycle. We found a sharp distribution of photon arrivals at each gate, which can be fitted well with a Gaussian lineshape with a full width at half maximum (FWHM) of 50~ps, as shown in Fig.~2(d). A significant gap between adjacent peaks prevents any ambiguity in assigning clock cycles to the received photons and hence the bit value.  In comparison, InGaAs APDs under conventional mode display much poorer timing performance (FWHM: 370~ps), as shown in Fig.~2(d). Moreover, adjacent photon peaks overlap at this photon arrival frequency, introducing ambiguity in assigning the clock cycle for photons received.  Typically, such a problem can be reduced by applying a small time window \cite{takesue07} to reject the most ambiguous photons, but such a solution not only reduces the detector efficiency but also adds to the complexity of the system. In light of such observations, we believe that SD APDs are preferable for applications, such as QKD, where photons arrive at regular intervals, even if improvements in InGaAs APD technology allow continuous operation in the future.

\section{Results and discussion}
We implemented the decoy protocol with three different intensity levels: $\mu$, $\nu_1$, $\nu_2$ ($\mu>\nu_1>\nu_2$), with duty cycles of 0.80, 0.16 and 0.04 respectively. The signal pulses ($\mu$) are used for key generation, while those of intensity $\nu_1$ and $\nu_2$ are used as decoy pulses to characterize the quantum channel. The most efficient decoy protocol requires a vacuum state ($\nu_2=0$). In the experiment, $\nu_2$ is set to be 29~dB lower than the signal pulse ($\mu$) due to the finite extinction ratio of the intensity modulator. Numerical simulations are used to determine signal and decoy intensities for an optimal secure key rate.

We first performed QKD across a fiber distance of 20~km, the most important distance for an optimal linear QKD network \cite{scarani08}.  Average photon fluxes of [$\mu$=0.55, $\nu_1$=0.10, $\nu_2$=7.6$\times$10$^{-4}$] were used. The high clock rate allows accumulation of sufficient photon events from a short session to account for statistical fluctuations.  During an uninterrupted session lasting 2.3~s, a total of $7.91\times10^6$ sifted signal bits were accumulated with a QBER of 2.53\%.  Based on parameters obtained experimentally, we are able to calculate single photon pulse characteristics using Eqns. (1 -- 3).  In order to account for the finite key size, we use statistical upper and lower bounds calculated at ten standard deviations for experimental parameters, to give a conservative bound on the key rate.  These parameters and results are summarized in Table~1. The achieved secure key rate of 1.02~Mbit/s is remarkable, being two orders of magnitude higher than the previous record of 10 kbit/s \cite{dynes07} for an unconditionally secure QKD protocol. We believe that this is the first time the secure key rate has exceeded 1~Mbit/s for any QKD protocol for this fiber length or longer.  Combined with the one-time pad cipher, such a key rate is sufficient to allow secure encryption of broadband communication, for example secure video links \cite{mink06}.

Figure 3 shows the fiber distance dependence of the raw and secure key rates, and the QBER. The same parameters as for the 20~km experiment are used for all the distances studied. Theoretical simulations of the key rates are also shown in Fig.~3. The measured QBER is dominated by the detector afterpulse noise \cite{yuan08} for fiber distances less than 60~km. The raw key rates as well as the secure key rates are in good agreement with the simulation. They decrease exponentially with fiber attenuation at 0.20~dB/km, apart from for short fibers ($\leq$20km). At 5.6~km, the secure key rate is 1.65~Mbit/s, noticeably less than expected, due to the count rate limitation imposed by the time-tagging electronics \cite{time-tagging}.  The secure key rates are determined to be 446 and 166~kbit/s for 40 and 60~km of fibers respectively.

For fiber lengths beyond 60~km, the detector dark count noise deteriorates the QKD performance, causing the secure key rate to fall faster than the fiber loss and eventually preventing key formation at a fiber distance of $\sim$110~km, as simulated in Fig.~3.  At 100.8~km of fiber, the QBER is measured to be 4.6\%, which is still sufficiently low for efficient key formation.  The raw key rate is reduced to 257~kbit/s due to the fiber attenuation, and correspondingly the QKD session time becomes noticeably longer at 16.5~s, in order to accumulate sufficient photon detection events to reduce statistical fluctuations. Including the effects of finite key length, a secure key rate of 10.1~kbit/s is achieved. This is a significant improvement over the previous records, where the key rate at an equivalent level of security is around $\sim$10~bit/s \cite{rosenberg07}.  The key rate is also more than 10 times greater than obtained in a recent high-speed QKD experiment \cite{tanaka08}, where a key rate of 0.78--0.82 kbit/s was estimated asymptotically using the decoy analysis method for a 97-km fiber link. The high key rate obtained here is a significant step towards the realization of long distance information-theoretically secure communication.

Finally, we would like to point out that there still remain challenges to overcome before realizing a complete gigahertz QKD system. These include remote synchronization, high speed error correction and privacy amplification, and high speed random number generation \cite{dynes08}. Gigahertz random number generators, which do not exist presently, are particularly important, as the security of the QKD relies upon Alice and Bob's ability to randomly modulate each optical pulse individually.

\section{Conclusion}
In summary, we have demonstrated high key rate QKD with gigahertz-clocked InGaAs APDs using the decoy protocol, which generates keys with unconditional security.  A significant leap in the secure key rate has been obtained, thanks to high performance SD detectors and the effectiveness of the decoy protocol.  With this advance, QKD is now practical for realizing high bandwidth information-theoretic secure communication. This is not only an exciting advance for cryptography, but also for the maturing field of quantum information science.

\begin{acknowledgments}
The authors would like to thank the EU for funding under the FP6 Integrated Project SECOQC.
\end{acknowledgments}

\newpage

\begin{table}
\caption{Summary of parameters and results for the 20-km decoy QKD experiment.  10 standard deviations are also shown for $Q_\mu$, $Q_{\nu_1}$, $Q_{\nu_2}$ and $\varepsilon_\mu$. }
\begin{ruledtabular}
\begin{tabular}{lclc}
\textbf{Parameter} & \textbf{Value}  & \textbf{Parameter} & \textbf{Value} \\
\hline
Fiber length & 20.06~km & $Q_{\mu}$ & $(8.680\pm0.025)\times10^{-3}$\\
$\mu$&0.55&$Q_{\nu_1}$&$(1.970\pm0.025)\times10^{-3}$\\
$\nu_1$&0.10&$Q_{\nu_2}$&$(4.470\pm0.225)\times10^{-4}$\\
$\nu_2$&$7.5\times10^{-4}$&$\varepsilon_\mu$&$(2.530\pm0.009)\%$\\
Prob.($\mu:\nu_1:\nu_2$)& 0.80:0.16:0.04 & $Q_1^L$ & $4.81\times10^{-3}$\\
$t$& 2.3~s & $\varepsilon_1^L$ & 2.10\% \\
$f_{EC}$& 1.10 & $R_{Secure}$ & 1.02~Mbit/s\\
\end{tabular}
\end{ruledtabular}
\end{table}

~~~~~~~

\newpage

\begin{figure}[h]
\centering\includegraphics[width=16cm]{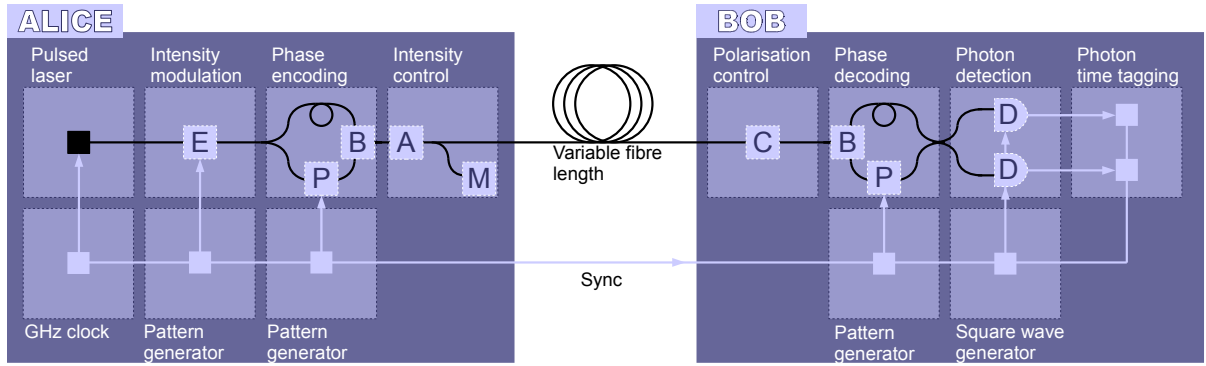}
\caption{Schematic of QKD system. [P] denotes a phase modulator, [B] a polarising beam splitter/combiner, [A] an attenuator, [M] an optical power meter, [C] a polarization controller and [D] InGaAs APD. All the modulating optics are driven by stand-alone gigahertz pattern generators with pre-loaded pseudo-random patterns.}
\end{figure}

\newpage
\begin{figure}[t]
\centering\includegraphics[width=12cm]{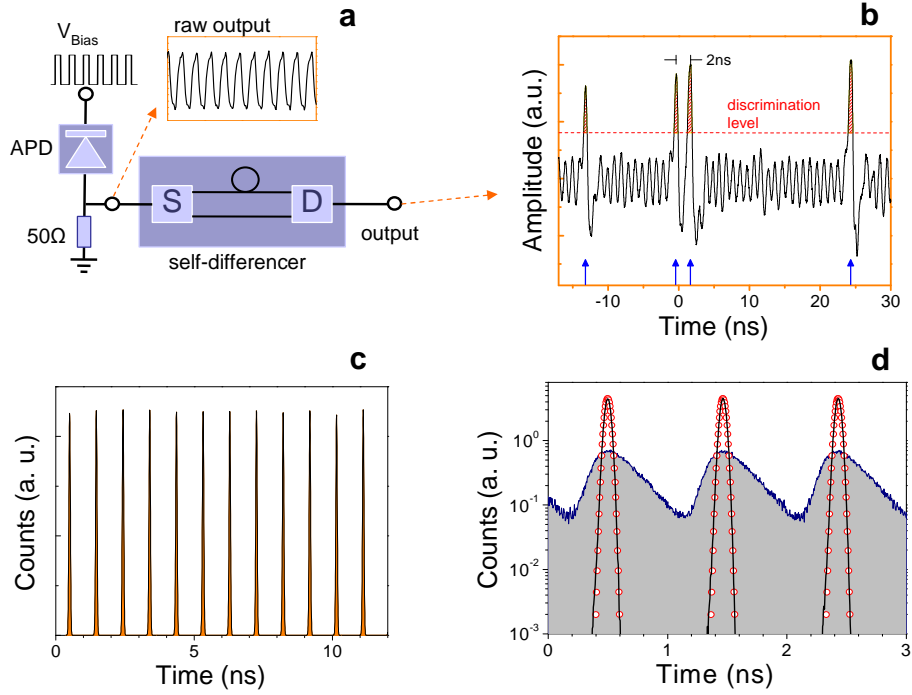}
\caption{Self-difference mode operation of an InGaAs APD. (a) Schematic for an APD in self-differencing mode. [S] is a 50:50 electrical signal splitter, [D] a subtractive combiner;
Co-axial cables connecting the splitter and combiner are made with precise lengths so that the time delay between them corresponds to one APD gating clock period. The APD raw output signal is dominated by the periodic capacitive response, as shown in the inset.  (b) A SD output recorded by an oscilloscope. The arrows indicate well discriminated avalanches.  (c) Histogram of photon arrival times obtained using the output  of the SD circuit.  (d) A comparison of histograms of photon arrivals obtained with the same APD under conventional gated Geiger mode operation (blue line) or SD mode (black line). For both measurements, the photon detection efficiency was set to 10\%. The red circles represent a Gaussian fit for histograms obtained with the SD circuit, giving a full width at half maximum of 50~ps.}
\end{figure}

\newpage
\begin{figure}[t]
\centering\includegraphics[width=12cm]{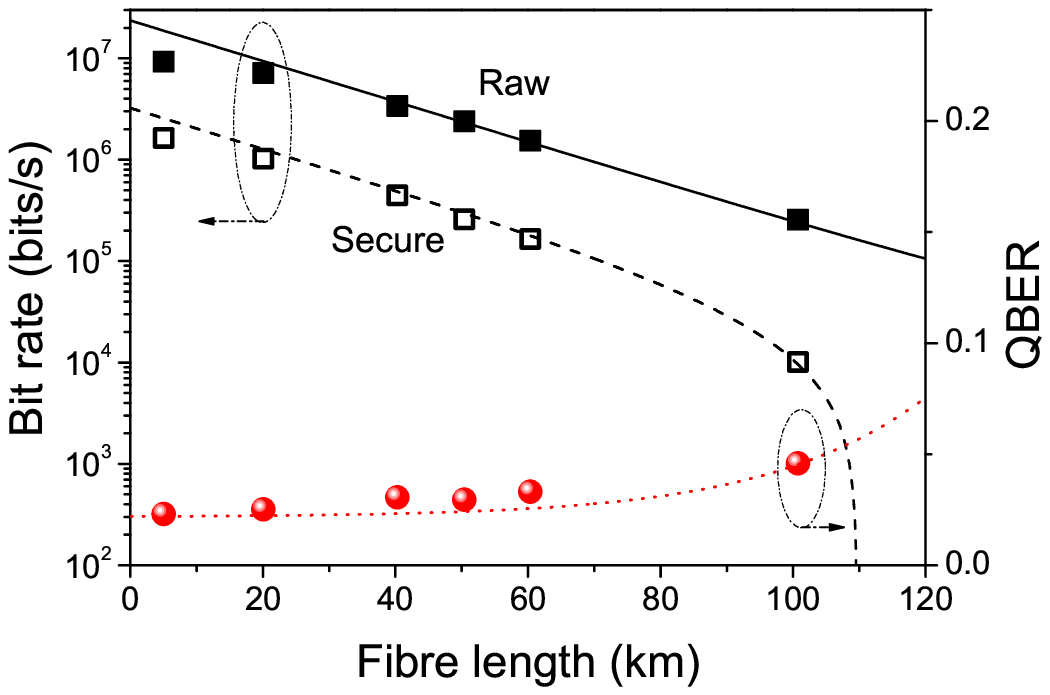}
\caption{Decoy QKD experimental results. Raw (filled squares),  secure (open squares) key rate, and the QBER (solid circles). Theoretical simulations are also shown for raw (solid lines), secure (dashed lines) key rates and the QBER (dotted line).  In simulations, Bob's detection efficiency is set to 5\% (corresponding 10\% of detector efficiency), detector dark count rate of $6.8\times10^{-6}$ and afterpulse rate 4.7\%, and a QBER of 0.3\% due to optical misalignment. All simulation parameters used are consistent with experimental results.}
\end{figure}


\begin{thebibliography} {99}
\bibitem{time-tagging}
Our time-tagging electronics record each photon with a 32-bit number to represent its arrival time. Due to restriction in the data bandwidth at around 160 Mbit/s per channel, for transferring photon information into computer memory, each channel can cope with only 5 million photons per second at most.
\end{thebibliography}
\end{document}